\documentclass[%
 reprint,
superscriptaddress,
 amsmath,amssymb,
aps,
prb,
floatfix,
]{revtex4-2}
\usepackage{silence}
\usepackage{lmodern}
\usepackage{siunitx}
\usepackage{chemfig}
\usepackage{amsmath}
\usepackage{amssymb}
\usepackage{color}
\usepackage{times}
\usepackage{epsf}
\usepackage{graphicx}
\usepackage{epsfig}
\usepackage{epstopdf}
\usepackage{dcolumn}
\usepackage{bm}
\usepackage{float}
\usepackage{url}
\usepackage{CJK}
\usepackage{ulem}
\usepackage{makecell}
\usepackage{newtxtext, newtxmath}
\usepackage{multirow}
\usepackage{subfigure}
\usepackage[colorlinks=true,linkcolor=blue,citecolor=blue,urlcolor=blue]{hyperref}

\def\<{\langle}
\def\>{\rangle}
\def\({\left(}
\def\){\right)}
\def\[{\left[}
\def\]{\right]}

\begin{document}
\title{Origin of Giant Phonon Magnetic Moment in Orbital Seebeck Effect:\\
a Heisenberg-type L-L Coupling}

\author{Hong Sun}
\thanks{These authors contribute equally to this work.}
\affiliation{Ministry of Education Key Laboratory of NSLSCS, Phonon Engineering Research Center of Jiangsu Province, Center for Quantum Transport and Thermal Energy Science, Institute of Physics Frontiers and Interdisciplinary
Sciences, School of Physics and Technology, Nanjing Normal University, Nanjing 210023, China}

\author{Jinxin Zhong}
\thanks{These authors contribute equally to this work.}
\affiliation{School of Energy and Materials, Shanghai Polytechnic University, Shanghai 201209, China}

\author{Yimin Yao}
\affiliation{State Key Laboratory of Materials for Integrated Circuits, Shenzhen Institute of Advanced Electronic Materials, Shenzhen Institutes of Advanced Technology, Chinese Academy of Sciences, Shenzhen 518055, China}

\author{Jun Zhou}
\email{zhoujunzhou@njnu.edu.cn}
\affiliation{Ministry of Education Key Laboratory of NSLSCS, Phonon Engineering Research Center of Jiangsu Province, Center for Quantum Transport and Thermal Energy Science, Institute of Physics Frontiers and Interdisciplinary Sciences, School of Physics and Technology, Nanjing Normal University, Nanjing 210023, China}

\author{Lifa Zhang}
\email{phyzlf@njnu.edu.cn}
\affiliation{Ministry of Education Key Laboratory of NSLSCS, Phonon Engineering Research Center of Jiangsu Province, Center for Quantum Transport and Thermal Energy Science, Institute of Physics Frontiers and Interdisciplinary Sciences, School of Physics and Technology, Nanjing Normal University, Nanjing 210023, China}

\date{\today}
\begin{abstract}
Inspired by the recent observation of the orbital Seebeck effect in $\alpha$-quartz, we identify an intrinsic amplification mechanism for thermally generated phonon angular momentum and phonon magnetic moment in chiral insulators. We propose a Heisenberg-type long-range coupling between phonon angular momenta, referred to here as L-L coupling, which opens a self-consistent feedback channel and strongly enhances the bare thermal response within linear response. Our calculations reveal a pronounced temperature- and size-dependent amplification, dominated by the off-diagonal channel, with the total phonon angular momentum enhanced by up to nearly two orders of magnitude as the system approaches the threshold from below. These findings suggest that L-L coupling may provide a microscopic origin of giant phonon magnetic moment the recently observed orbital Seebeck effect in $\alpha$-quartz.
\end{abstract}

\maketitle
\textit{Introduction}\textemdash The generation and manipulation of magnetism in various magnetic materials was carried out primarily by electrical and optical means \cite{RevModPhys.82.2731,siegrist2019light,doi:10.1021/acsnano.3c13002}. Recently, axial phonons \cite{PhysRevLett.115.115502,juraschek2025chiral}, lattice excitations carrying phonon angular momentum (PAM) \cite{PhysRevLett.112.085503}, have emerged as an alternative route for generating and controlling magnetic responses in solids without magnetic ions. This development stems from the recognition that phonons can possess not only effective charge \cite{PhysRev.173.833} but also angular momentum, and therefore a finite magnetic moment \cite{PhysRev.6.239,PhysRevMaterials.1.014401,PhysRevMaterials.3.064405}. As a result, the interplay between phonon and magnetism leads to a growing body of experiments, including the phonon Zeeman effect \cite{doi:10.1021/acs.nanolett.0c01983,PhysRevLett.128.075901,doi:10.1126/sciadv.adj4074,PhysRevB.109.155426,doi:10.1073/pnas.2304360121}, light-induced phonon magnetization \cite{nova2017effective,PhysRevResearch.4.013129,doi:10.1126/science.adi9601,basini2024terahertz}, and phonon-driven magnetic switching \cite{afanasiev2021ultrafast,sasaki2021magnetization,doi:10.1126/sciadv.abm4005,davies2024phononic,PhysRevResearch.5.L022039,doi:10.1021/acs.jpclett.5c00304,PhysRevB.111.134414}. These experiments provided compelling evidence that such axial phonon-magnetic effects can be substantial across a wide range of materials. Nevertheless, the microscopic origin of the phonon magnetic moment remains under active debate, particularly because of recently observed orbital Seebeck effect (OSE) in $\alpha$-quartz \cite{nabei2026orbital}. The mechanism of the giant phonon magnetic moment induced by the temperature gradient is not yet fully understood.

\begin{figure}[htbp]
\centering
\includegraphics[width=0.90\linewidth]{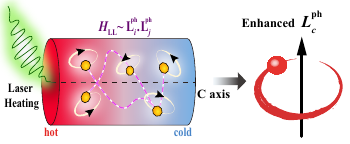}
\caption{Schematic illustration of the self-consistent amplification mechanism for thermally generated PAM in a symmetric-top-like chiral insulator. Laser heating establishes an axial temperature gradient along the principal $\mathrm{c}$ axis of the sample, which induces local PAM. Through L-L coupling, local rotational motions become correlated and feed back positively on response coefficient, leading to an enhanced PAM.}
\label{fig1}
\end{figure}

\par
To date, several microscopic scenarios have been proposed to account for these unusually large axial phonon-magnetic responses. Within a purely ionic picture, the phonon magnetic moment is expected to be limited by the ionic mass scale and thus remain on the order of the nuclear magneton, far smaller than the values observed in experiments. This discrepancy has motivated a variety of theories in which PAM is transferred to or coupled with electronic degrees of freedom, thereby enabling a much larger effective magnetic moment. Existing proposals include perturbative electron-phonon mechanisms \cite{PhysRevLett.133.266702,PhysRevB.110.094401,yao2026dynamicalorbitalangularmomentum,sato2025orbitalaccumulationinducedchiral}, non-adiabatic \cite{tpjd-dh1m} or Berry-phase formulations \cite{PhysRevB.86.104305,PhysRevLett.127.186403,6m7v-p99w}, Floquet descriptions of driven phonons \cite{shin2018phonon}, and effective gauge-field \cite{PhysRevLett.127.125901,chen2025geometricoriginphononmagnetic,chaudhary2025chernsimonstypecrosscorrelationsgeometric} or inertial pictures \cite{PhysRevB.108.064510,PhysRevResearch.4.L012004,kp6b-2kxz}. Despite their different formulations, these approaches pointed to a common conclusion: the electron-phonon interaction is essential. However, the recent observation in $\alpha$-quartz, a prototypical wide-gap insulator, challenges this "common conclusion", because explanations relying on electronic enhancement of the phonon magnetic moment become less natural in such an insulator. This tension between existing theory and experiment therefore requires a purely phononic mechanism capable of generating giant phonon magnetic moment in chiral insulating systems.
\par
In this Letter, we propose a purely phononic mechanism, namely a Heisenberg-type L-L coupling between PAM, hereafter referred to as L-L coupling. Rooted in Landau’s description of vibration and rotation \cite{landau1958quantum}, this coupling arises as an intrinsic feature of symmetric top-like systems and captures how local rotational dynamics induced by lattice vibrations. Starting from this coupling, within a linear-response framework, we derive a self-consistent expression for thermally generated PAM. We then apply this formalism to $\alpha$-quartz and show from first-principles calculations that the interplay of crystal symmetry, phonon texture, and L-L feedback gives rise to a pronounced temperature- and size-dependent enhancement of the thermal response, which can be experimentally detected.
\par
We consider a solid body with symmetric-top-like structure with principal moments of inertia satisfying $I_{\text{A}}=I_{\text{B}}\neq I_{\mathrm{C}}$. 
A microscopic description should retain not only the vibrational motion of the lattice but also the rotational degree of freedom implied by the symmetric-top-like geometry. This naturally leads to Landau’s coupled vibration-rotation framework \cite{landau1958quantum}, in which the total angular momentum is decomposed into two parts:
\begin{equation}
\mathbf{L}=\sum_{i}\[
\mathbf{R}_{i}\times\(m_{i}\dot{\mathbf{u}}_{i}\)
+\mathbf{u}_{i}\times\(m_{i}\dot{\mathbf{u}}_{i}\)
\].
\label{angular}
\end{equation}
Here $\mathbf{R}_{i}$ and $\mathbf{u}_{i}$ are the equilibrium position and displacement of the $i$-th atom, and $m_{i}$ is its mass. Here we neglect the terms containing $\dot{\mathbf{R}_{i}}$ because we ignore macroscopic rotation. It is also convenient to choose the center of mass as the origin of the coordinates.
Then the first term on the right side of Eq. (\ref{angular}), noted as $\mathbf{L}^{\mathrm{rot}}$, is the angular-momentum contribution arising from the vibrations evaluated with respect to the origin of coordinate. This term describes the micro-rotation of the body. The second term, noted as $\mathbf{L}^{\mathrm{ph}}$, originates from the locally circular lattice motion around their equilibrium positions. Consequently, expressed in the principal-axis frame, the rotational Hamiltonian naturally separates into three parts: a bare micro-rotational term ${H}_{\mathrm{rot}}$ in Eq. (\ref{hamiltonian}a), a Coriolis-like coupling between vibration and micro-rotation ${H}_{\mathrm{cor}}$ in Eq. (\ref{hamiltonian}b), and a term which describes the coupling between PAM (noted as L-L coupling $H_{\mathrm {LL}}$) of different atoms in Eq. (\ref{hamiltonian}c),
\begin{subequations}
\begin{align}
H_{\mathrm{rot}}
&=\frac{1}{2I_{\mathrm A}}{\mathbf L}^{2}
+\frac{1}{2}\(\frac{1}{I_{\mathrm C}}-\frac{1}{I_{\mathrm A}}\)L^{2}_{c},\\
H_{\mathrm{cor}}
&=\frac{1}{I_{\mathrm A}}\mathbf{L}\cdot\mathbf{L}^{\mathrm {ph}}
+\(\frac{1}{I_{\mathrm C}}-\frac{1}{I_{\mathrm A}}\)L_{c}L^{\mathrm {ph}}_{c},\\
H_{\mathrm{LL}}&=\frac{1}{2I_{\mathrm {A}}}\(\mathbf{L}^{\mathrm {ph}}\)^{2}
+\frac{1}{2}\(\frac{1}{I_{\mathrm C}}-\frac{1}{I_{\mathrm A}}\)\(L^{\mathrm {ph}}_{c}\)^{2}.
\end{align}
\label{hamiltonian}
\end{subequations}

Figure (\ref{fig1}a) shows that $H_{\mathrm{LL}}$ is a Heisenberg-type coupling between arbitrary different atoms, $i$ and $j$. Because it contains $\sum_{i,j}\mathbf{L}^{\mathrm{ph}}_{i}\cdot \mathbf{L}^{\mathrm{ph}}_{j}$. Such mathematical form is analogous to the traditional Heisenberg Hamiltonian $\sim\sum_{i,j}\mathbf{S}_{i}\cdot\mathbf{S}_{j}$. 
Physically, there are three differences between $H_{\mathrm{LL}}$ and the Heisenberg Hamiltonian: 1) $\mathbf{L}^{\mathrm{ph}}_{i}$ and $\mathbf{L}^{\mathrm{ph}}_{j}$ are usually zero in absence of external field, spins $\mathbf{S}_{i}$ and $\mathbf{S}_{j}$ are intrinsic and nonzero in magnets. 2) The coupling parameter between $\mathbf{L}^{\mathrm{ph}}_{i}$ and $\mathbf{L}^{\mathrm{ph}}_{j}$ is determined by moment of inertia which is always positive, in other words $\mathbf{L}^{\mathrm{ph}}_{i}$ prefers to be parallel to $\mathbf{L}^{\mathrm{ph}}_{j}$. In contrast, the exchange parameter in Heisenberg Hamiltonian could be positive or negative. 3) The L-L coupling is not only short-range but also long-range.

\par
We point out that, ${\bf L}$ and $L_{c}$ in Eqs. (\ref{hamiltonian}a) and (\ref{hamiltonian}b) are set to be zero, $H_{\mathrm{LL}}$ is the origin of giant phonon magnetic moment in chiral materials observed by the OSE experiment \cite{nabei2026orbital}. 
Applying a temperature gradient along $c$-direction $\frac{\partial T}{\partial x_{c}}$, phonon distribution can be driven out of equilibrium and generates an imbalance among angular-momentum-carrying phonon modes, thereby producing a net PAM as shown in Fig. (\ref{fig1}). If the L-L coupling is not considered, one can easily calculate the average PAM density along $c$-direction $\<L^{\text{ph}}_{c}\>^{\mathrm{NL}}$ by using the Boltzmann transport equation \cite{PhysRevLett.121.175301,PhysRevB.106.094311}:
\begin{equation}\label{eq3}
\<L^{\text{ph}}_{c}\>^{\mathrm{NL}}=
-\frac{\tau}{V}\sum_{\mathbf{k},\sigma}l^{c}_{\mathbf{k}\sigma}v^{c}_{\mathbf{k}\sigma}
\frac{\partial f^{0}_{\mathbf k\sigma}}{\partial T}
\frac{\partial T}{\partial x_{c}}
\equiv \gamma^{\mathrm{NL}} \frac{\partial T}{\partial x_{c}}.
\end{equation}
Here $\gamma^{\mathrm{NL}}$ is the response coefficient determined by phonon relaxation time $\tau$, phonon group velocity $v^{c}_{\mathbf{k}\sigma}$ and phonon polarization $l^{c}_{\mathbf{k}\sigma}$ with phonon wave vector ${\mathbf k}$ and phonon mode $\sigma$. $V$ is the volume and $f^{0}_{\mathbf k\sigma}$ is the Bose-Einstein distribution function. 

Once L-L coupling is considered, each $\mathbf{L}^{\mathrm{ph}}_{i}$ strongly interact with other PAM via Eq. (\ref{hamiltonian}c). Using mean field approximation, $\mathbf{L}^{\mathrm{ph}}_{i}$ feels a mean field $\sim\sum_{j}\mathbf{L}^{\mathrm{ph}}_{j}$. The $c$-component of this field is proportional to the total PAM density $\<L^{\text{ph}}_{c}\>^{\mathrm{total}}$ which is unknown. Because it can be fed back into the system through L-L coupling, giving rise to an additional contribution proportional to the total response,
\begin{equation}
\<L^{\text{ph}}_{c}\>^{\mathrm{LL}}=\gamma^{\mathrm{LL}}\<L^{\text{ph}}_{c}\>^{\mathrm{total}}.
\label{eq4}
\end{equation}
The coefficient $\gamma^{\mathrm{LL}}$ contains both diagonal $\gamma^{\mathrm{LL}}_{\mathrm{D}}$ and off-diagonal $\gamma^{\mathrm{LL}}_{\mathrm{OD}}$ contributions of the PAM operator, and its detailed form is given in the Supplemental Material (SM).
It is straightforward to find that the total PAM density remains proportional to the temperature gradient, we obtain a self-consistent relation
\begin{equation}
\<L^{\text{ph}}_{c}\>^{\mathrm{total}}=\alpha\frac{\partial T}{\partial x_{c}}=
\gamma^{\mathrm{NL}}\frac{\partial T}{\partial x_{c}}+
\gamma^{\mathrm{LL}}\alpha\frac{\partial T}{\partial x_{c}}.
\label{eq5}
\end{equation}
Combing Eqs. (\ref{eq4}) and (\ref{eq5}), we obtain
\begin{equation}\label{eq7}
\< L^{\mathrm{ph}}_{c}\>^{\mathrm{total}}=\frac{\gamma^{\mathrm{NL}}}{1-\gamma^{\mathrm{LL}}}\frac{\partial T}{\partial x_{c}}.
\end{equation}
This result shows a clear amplification effect with factor $\eta=1/(1-\gamma^{\mathrm{LL}})$. The bare thermal driving is encoded in $\gamma^{\mathrm{NL}}$, whereas the effect of L-L coupling enters through amplification factor. The detailed derivation of $\gamma^{\mathrm{LL}}$ and $\gamma^{\mathrm{NL}}$ is carried out by using the Kubo formula in SM. 

\begin{figure*}[htbp]
\centering
\includegraphics[width=0.99\linewidth]{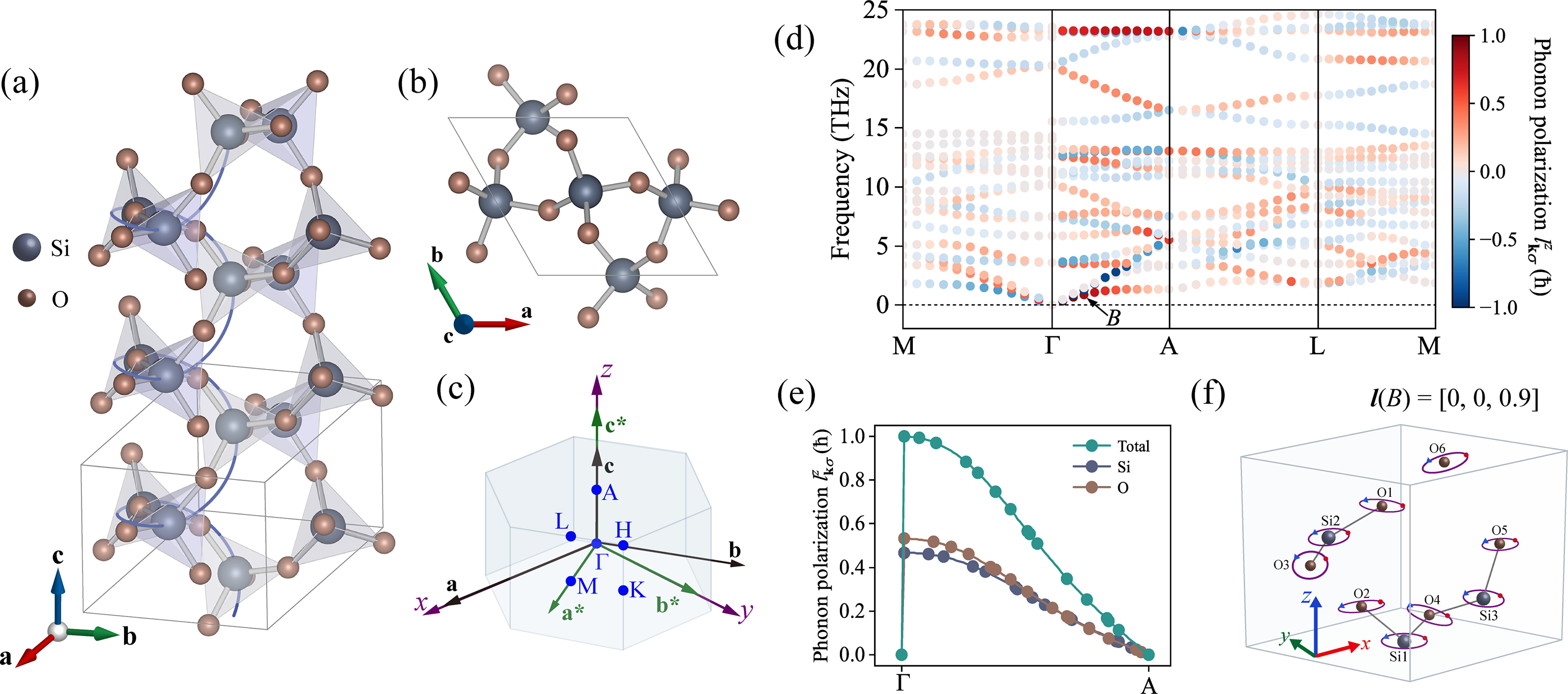}
\caption{(a) Crystal structure of $\alpha$-quartz, highlighting the helical arrangement of SiO$_{4}$ tetrahedra along the $c$-axis. (b) Corresponding top view of the quartz structure. (c) First Brillouin zone and the high-symmetry points (blue). The relations among the real-space lattice vectors ($\mathbf{a,b,c}$; black), reciprocal-space lattice vectors ($\mathbf{a^{*},b^{*},c^{*}}$; green), and Cartesian axes ($x,y,z$; purple) are indicated. (d) Phonon dispersion color coded by phonon polarization $l_{\mathbf{k}\sigma}^{z}$. (e) Atomic species-projected phonon polarization of the first acoustic branch along the $\Gamma$-$\mathrm{A}$ direction. (f) Real-space visualization of the atomic rotational motion for the phonon mode at point $B$ marked in (d). The initial phase (red dots), motion trajectories (purple curves), and motion directions (blue triangles) are indicated.}
\label{fig2}
\end{figure*}
\par

We now choose $\alpha$-quartz to exemplify our theory. $\alpha$-quartz is a chiral material with primitive unit cell containing nine atoms and is composed of corner-sharing SiO$_4$ tetrahedra as shown in Fig.~\ref{fig2}(a). Corner-sharing tetrahedra are arranged into a helical network along the $c$-axis, forming a right-handed spiral framework in real space. The corresponding top view in Fig.~\ref{fig2}(b) further reveals the projected in-plane arrangement of this framework, in which three equivalent branches are related by 120$^\circ$ rotation, making explicit the threefold rotational symmetry about the $c$-axis. Together with the helical stacking along the same axis, this reflects the characteristic threefold screw symmetry of quartz. 

\begin{figure*}[htbp]
\centering
\includegraphics[width=0.94\linewidth]{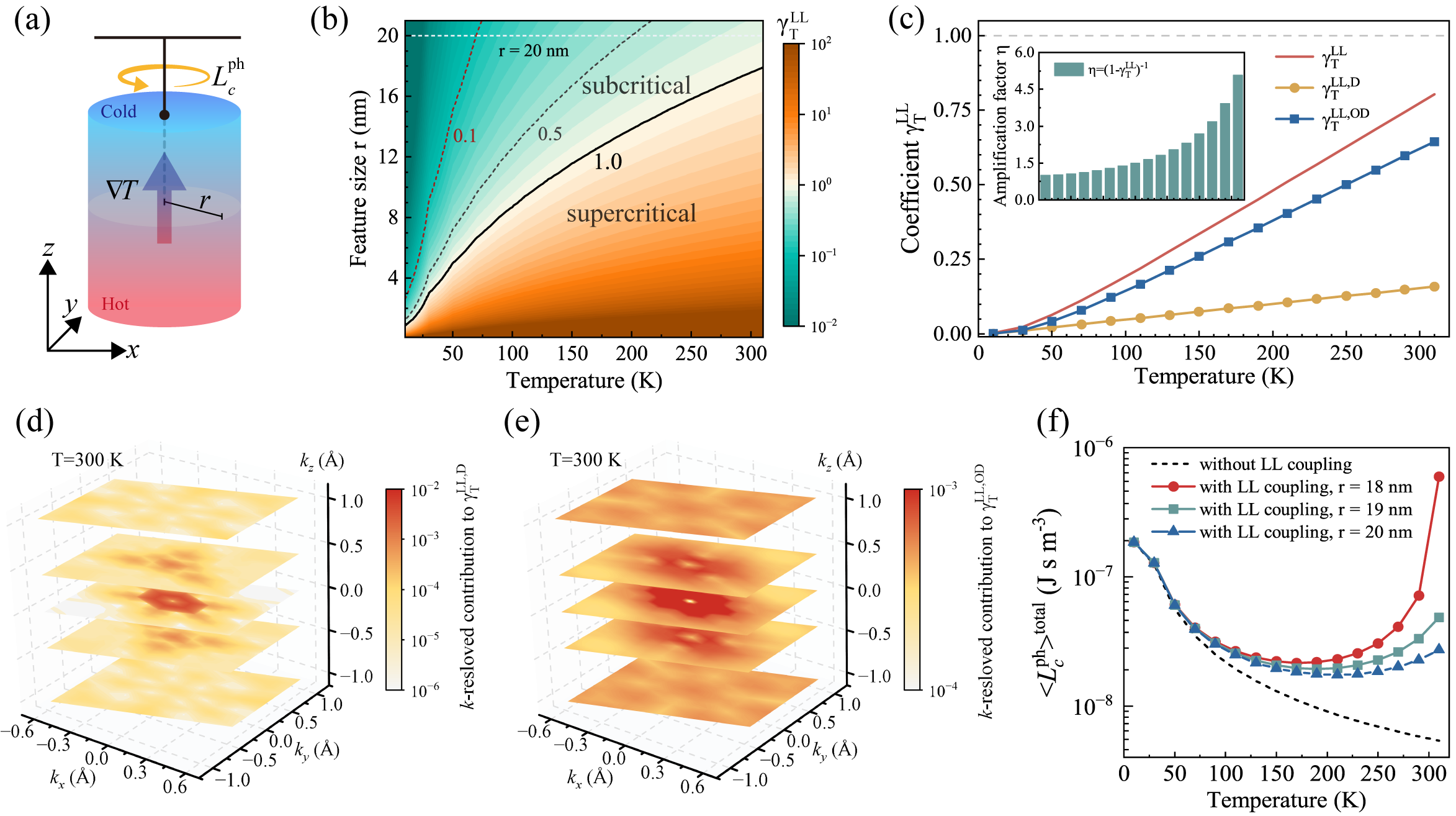}
\caption{(a) Schematic illustration of the cylindrical sample under an axial temperature gradient $\nabla T$, which induces a PAM. (b) Temperature-size phase diagram of the L-L coefficient $\gamma^{\mathrm{LL}}$. The contours at 0.1, 0.5, and 1.0 are shown, with the $\gamma^{\mathrm{LL}} = 1$ boundary separating the subcritical and supercritical regimes. The horizontal dashed line marks the representative cut at $r=20$ nm. (c) Temperature dependence of $\gamma^{\mathrm{LL}}$ at 20 nm, together with its diagonal and off-diagonal contributions. The inset shows the corresponding amplification factor $\eta=(1-\gamma^{\mathrm{LL}})^{-1}$. (d, e) Brillouin-zone-resolved contributions to the diagonal and off-diagonal parts at $T=300$ K and $r=20$ nm, shown on selected $k_{z}$ slices. The color scales are chosen independently for clarity. (f) Total PAM density as a function of temperature without and with LL coupling for different cylinder radii, showing the strong size-dependent enhancement}
\label{fig3}
\end{figure*}
\par
In our calculations, it is convenient to fix $c$-axis to $z$-direction and the first Brillouin zone with selected high-symmetry points are plotted in Fig.~\ref{fig2}(c). Fig.~\ref{fig2}(d) presents the phonon dispersion color coded by the phonon polarization which characterizes the PAM for given $\mathbf{k}$ and $\sigma$. Since we are concerned with angular-momentum excitations along the $c$-axis, only the $z$-component, $l^{z}_{\mathbf{k}\sigma}$, is shown. For clarity, we further restrict the plot to the high-symmetry paths that exhibit pronounced phonon polarization features; full first-principles calculation details and the complete phonon dispersion are provided in the SM. As can be seen, finite phonon polarization is distributed nonuniformly among the 27 branches. Starting from the $\Gamma$ point, the phonon polarization gradually develops along selected directions, but vanishes again at the high-symmetry points $\mathrm M$ and $\mathrm L$. This disappearance can be traced to the little-group symmetry at these points: a twofold rotation about an in-plane axis leaves these wave vectors invariant modulo a wave vector while reversing the $z$-component of PAM, thereby forcing $l_{\mathbf{k}\sigma}^{z}=0$. By contrast, along the $\Gamma$-$\mathrm{A}$ direction, which follows the screw axis of the crystal, finite $l_{\mathbf{k}\sigma}^{z}$ is symmetry-allowed and becomes especially pronounced. This makes the $\Gamma$-$\mathrm{A}$ path the clearest momentum-space manifestation of the chiral symmetry of quartz and highlights a strong path selectivity in PAM texture.
\par
Fig.~\ref{fig2}(e) zooms in on the lowest acoustic branch along the $\Gamma$-$\mathrm{A}$ direction. A striking feature is that, although the phonon polarization vanishes exactly at $\Gamma$ point, it rises abruptly to an almost fully polarized value immediately away from the zone center and then gradually decreases toward zero as the wave vector approaches $\mathrm{A}$. This sharp onset is a direct manifestation of acoustical activity in a chiral crystals \cite{PhysRevB.2.2049}. As predicted by the first-order-in-$k$ theory \cite{PhysRev.170.673}, an infinitesimal wave vector along the $c$-axis activates a chiral splitting of the transverse acoustic pair and turns the linearly polarized modes into circularly polarized states, causing $l^{z}_{\mathbf{k}\sigma}$ to emerge almost discontinuously from the zone center.
\par
Beyond this sharp onset, the persistence of a large polarization along the branch can be traced to the atomic decomposition: the Si and O atoms contribute constructively and with comparable magnitudes, thereby sustaining a large total phonon polarization over a broad portion of the path. This picture is further corroborated by the real-space vibrational pattern shown in Fig.~\ref{fig2}(f), which visualizes the atomic motion for the representative mode at point $B(\mathbf{k}_{B}=0.091\times \mathbf{c}^{*})$ marked in Fig.~\ref{fig2}(d), where the phonon polarization is already large, $\textbf{\textit l}_{B}=[0,0,0.9]$. 
Both Si and O ions execute elliptical trajectories with the same projected sense of rotation about the $c$ axis, yielding a cooperative rather than compensating contribution to $l_{\mathbf{k}\sigma}^{z}$. 
\par

Fig.~\ref{fig3}(a) illustrates the setup we considered: a cylindrical quartz sample of radius $r$ subject to an axial temperature gradient. In this geometry, the thermal bias induces a PAM along $z$-direction, as allowed by the crystal symmetry. The moment of inertia is $I_{\mathrm{c}}=\frac{1}{2}\rho V r^{2}$ with mass density $\rho$. Fig.~\ref{fig3}(b) shows the calculated temperature- and size-dependence of $\gamma^{\mathrm{LL}}$. In the low-temperature and large-size region, $\gamma^{\mathrm{LL}}$ remains far below unity, so the L-L feedback remains weak and the system stays deeply in the subcritical regime. With increasing temperature, however, the coefficient grows steadily and the contours shift upward, indicating that the feedback becomes effective for progressively larger feature sizes. At fixed temperature, reducing $r$ produces the same trend even more strongly, eventually driving the system across the $\gamma^{\mathrm{LL}}= 1$ threshold, which means $\eta \rightarrow \infty$. This contour therefore separates the parameter space into subcritical and supercritical regimes. For $\gamma^{\mathrm{LL}}<1$, the self-consistent amplification remains finite within the linear theory. By contrast, once $\gamma^{\mathrm{LL}}>1$, the weak-response solution assumed in the linearized treatment becomes unstable against L-L feedback. In this sense, the supercritical region marks the breakdown of the weak-response solution rather than a directly evaluated physical steady state. In the present work, we therefore do not attempt to characterize the response beyond this threshold, and instead restrict our analysis to the subcritical regime and to the approach to the threshold from below.
\par
Fig.~\ref{fig3}(c) shows the corresponding temperature dependence of $\gamma^{\mathrm{LL}}$ when $r=20$ nm as marked by horizontal dash line in Fig.~\ref{fig3}(b). As the temperature increases, $\gamma^{\mathrm{LL}}$ rises monotonically, indicating a continuous strengthening of the L-L feedback while the system remains on the subcritical side throughout the range considered. The decomposition further shows that this growth is dominated by the off-diagonal contribution, whereas the diagonal part remains much smaller throughout, identifying the off-diagonal channel as the primary source of the L-L feedback buildup. Moreover, the inset displays the corresponding amplification factor $\eta$, showing that the attainable enhancement increases from nearly unity at low temperature to about fivefold near room temperature.
\par
Figs.~\ref{fig3}(d) and (e) further show the Brillouin-zone-resolved contribution at $T=300$ K. The diagonal part is seen to be highly localized in momentum space, with appreciable weight confined to a narrow region around the central slices. In contrast, the off-diagonal part spreads over a much broader region and persists across multiple $k_{z}$ planes. This diffuse interband contribution originates from the dense multibranch phonon structure throughout the Brillouin zone, which allows interbranch coupling to remain active over an extended momentum-space region. The difference is therefore governed less by the peak magnitude at individual $\mathbf{k}$ points than by the much larger phase space over which the off-diagonal contribution remains finite, ultimately making this channel dominate the total L-L feedback.
\par
Finally, we assess the actual enhancement of thermally generated PAM by the L-L coupling. Fig.~\ref{fig3}(f) displays the temperature dependence of the total PAM density with and without L-L coupling for different cylinder radii. At low temperatures, all curves nearly overlap, indicating that the L-L feedback is negligible in this regime. As temperature increases, curves with L-L coupling progressively deviate upward from the no-coupling baseline, and the deviation becomes increasingly sensitive to the sample size. Near room temperature, the total PAM density increases from roughly $10^{-9}$-$10^{-8}$ \si{\joule\second\per\meter\cubed} without L-L coupling to approximately $10^{-8}$-$10^{-7}$ \si{\joule\second\per\meter\cubed} with L-L coupling, reaching about \SI{5e-7}{\joule\second\per\meter\cubed} for $r=18$ nm, corresponding to an enhancement of up to nearly two orders of magnitude. Such a substantial enhancement suggests that the effect could be within experimental reach through the resulting rigid-body rotational response of the crystal, for example via torsional or torque-based measurements \cite{zhang2025measurement}.
\par
From the magnetic perspective, this enhanced angular momentum should also be accompanied by a correspondingly enhanced phonon magnetic response. Because the Born effective charges in quartz are on the order of a few elementary charges, a simple charge-to-mass estimate is sufficient to place the associated phonon magnetic moment in the nuclear-magneton scale \cite{PhysRevMaterials.3.064405}. Using the standard relation $\mu_{\mathrm{ph}}=\gamma\hbar$, with $\gamma$ determined at the order-of-magnitude level by the ionic charge-to-mass ratio, one can estimate $\mu_{\mathrm{ph}}$ to range from roughly 
$10^{-2}\mu_{\mathrm{N}}$ to $10^{-1}\mu_{\mathrm{N}}$ or equivalently from about $10^{-5}\mu_{\mathrm{B}}$ to $10^{-4}\mu_{\mathrm{B}}$ per phonon. Combined with the enhanced angular-momentum density in Fig.~\ref{fig3}(f), this corresponds to a magnetic response of order 1-10 A m$^{-1}$, or an equivalent magnetic-field scale of a few $\mu$T, suggesting that the associated signal could also be within experimental reach. We further pointed out that, when an external magnetic field is applied, the phonon magnetic moment would be further enhanced.
\par
In summary, we have shown that thermally generated PAM in a chiral insulator can be strongly amplified through intrinsic lattice feedback. The central result is that a Heisenberg-type long-range coupling between PAM opens a self-consistent amplification channel, allowing a weak bare thermal response to evolve into a much larger observable effect. In $\alpha$-quartz, this mechanism is rooted microscopically in the screw-symmetry-controlled phonon texture and manifests macroscopically as a pronounced temperature- and size-dependent enhancement of the total angular momentum, with the buildup of the L-L feedback governed primarily by the off-diagonal channel. Our results therefore shift the emphasis from bare thermal generation alone to the importance of feedback-driven amplification, and suggest that L-L coupling may provide a microscopic origin of the recently observed orbital Seebeck effect in $\alpha$-quartz.
\begin{acknowledgments}

This work is supported by the National Key R\&D Program of China (No.2022YFA1203100).
\end{acknowledgments}
\normalem
\bibliography{ref}
\end{document}